\documentstyle[11pt,graphicx,twocolumn]{article}

\newcommand{\be}{\begin{equation}}
\newcommand{\ee}{\end{equation}}
\newcommand{\ba}{\begin{eqnarray}}
\newcommand{\ea}{\end{eqnarray}}

\input{tcilatex}

\begin{document}

\title{{\bf Pre-formed Cooper pairs and Bose-Einstein condensation in cuprate
superconductors}}
\author{M. Casas$^{a}$, M. de Llano$^{b}$, A. Puente$^{a}$, A. Rigo$^{a}$ and M.A.
Sol\'{\i}s$^{c}$ \vspace{0.4cm} \\
$^{a}$Departament de F\'{\i}sica, Universitat de les Illes Balears,\\
07071 Palma de Mallorca, Spain \vspace{0.2cm}\\
$^{b}$Instituto de Investigaciones en Materiales, Universidad Nacional
Aut\'{o}noma de M\'{e}xico,\\
Apdo. Postal 70-360, 04510 M\'{e}xico, DF, Mexico \vspace{0.2cm}\\
$^{c}$Instituto de F\'{\i}sica, Universidad Nacional Aut\'onoma de M\'exico, 
\\
Apdo. Postal 20-364, 01000 M\'{e}xico, DF, Mexico.}
\maketitle

\begin{abstract}
A two-dimensional (2D) assembly of noninteracting, temperature-dependent,
pre-formed Cooper pairs in chemical/thermal equilibrium with unpaired
fermions is examined in a binary boson-fermion statistical model as the
Bose-Einstein condensation (BEC) singularity temperature $T_{c}$\ is
approached from above. \ Compared with BCS theory (which is {\it not} a BEC
theory) substantially higher $T_{c}$'s\ are obtained without any adjustable
parameters, that fall roughly within the range of empirical $T_{c}$'s for
quasi-2D cuprate superconductors. \newline
\end{abstract}

A possible interpretation of the ``pseudo-gap'' observed in some
superconductors {\it above }$T_{c}$ is that it arises \ simultaneously with
the formation of \ ``pre-formed'' Cooper pairs (CPs). We propose here that
such objects emerge naturally as the nonzero-total (or, -center-of-mass)
momentum (CMM)\ CPs that are entirely neglected in ordinary BCS theory.\ 

Consider a 2D system of N fermions of mass $m$ confined in a square of area $%
L^{2}$ and interacting pairwise via the BCS model interaction $V_{{\bf k},%
{\bf k}^{\prime }}=-V\;$when$\;\mu (T)-\hbar \omega _{D}<\epsilon
_{k_{1}}(\equiv \hbar ^{2}k_{1}^{2}/2m)$ and$\ \epsilon _{k_{2}}<\mu
(T)+\hbar \omega _{D}$, and zero otherwise, where ${\bf k}\equiv {\frac{1}{2}%
}({\bf k}_{1}-{\bf k}_{2})$ is the relative wavevector of the two particles; 
$V_{{\bf k},{\bf k}^{\prime }}$ the 2D double Fourier integral of the
underlying non-local interaction $V({\bf r},{\bf r}^{\prime })$ in the
relative coordinate ${\bf r}={\bf r}_{1}-{\bf r}_{2};\ \ \mu (T)$ the ideal
Fermi gas (IFG) chemical potential which at $T=0$ becomes the Fermi energy $%
E_{F}\equiv \hbar ^{2}k_{F}^{2}/2m$ with $k_{F}$ the Fermi wavenumber; $%
2\hbar \omega _{D}\equiv \hbar ^{2}k_{D}^{2}/m$ the energy width of the
annulus centered around the Fermi circle and where the interaction is
nonzero, with $\omega _{D}$ the Debye frequency. For $V>0$ this model
interaction mimics the net effect of an attractive electron-phonon
interaction overwhelming the interfermion Coulomb repulsions.

If $\hbar {\bf K}=\hbar ({\bf k}_{1}+{\bf k}_{2})$ is the CMM of a CP, let $%
E_{K}$ be its {\it total} energy (besides the CP rest-mass energy). The
original eigenvalue CP \cite{Coo} equation for a pair of fermions at $T=0$
immersed in a background of $N-2$ inert, spectator fermions lying within a
(sharp) Fermi circular perimeter of radius $k_{F}$, is then 
\begin{equation}
1=V{\sum_{{\bf k}}}^{^{\prime }}\frac{\theta (k_{1}-k_{F})\,\,\theta
(k_{2}-k_{F})}{2\epsilon _{k}+{\hbar ^{2}}K{^{2}/}4m-E_{K}},
\label{eq:cooper}
\end{equation}
where $\theta (x)$ is the Heaviside unit step function, and the prime on the
summation sign denotes the conditions $k_{1,2}\equiv |{%
{\frac12}%
}{\bf K}\pm {\bf k}|<(k_{F}^{2}+k_{D}^{2})^{1/2}\;$ensuring that our pair of
fermions {\it above} the Fermi ``surface'' cease interacting beyond the
annulus of energy width $2\hbar \omega _{D}$, thereby restricting the
summation over ${\bf k}$ for a given fixed ${\bf K}$. Without these
restrictions (\ref{eq:cooper}) would just be the Schr\"{o}dinger equation in
momentum space for the pair. Setting $E_{K}\equiv 2E_{F}-\Delta _{K}$, the
pair is {\it bound} if $\Delta _{K}>0$, and (\ref{eq:cooper}) becomes an
eigenvalue equation for the (positive) pair binding energy $\Delta _{K}$.
Our $\Delta _{K}$ and $\Delta _{0}$ should {\it not} be confused with the
BCS energy gap $\Delta (T)$ at $T=0$. \ Let $\lambda \equiv g(E_{F})V\geq 0$
be the usual BCS dimensionless coupling constant. Here $g(E_{F})$ is the
electronic density-of-states (for each spin) at the Fermi surface in the
normal (i.e., interactionless) state, which in 2D is $g(\epsilon
)=L^{2}m/2\pi \hbar ^{2}\equiv g$, a constant. \ The Cooper equation (\ref
{eq:cooper}) for the unknown quantity $\Delta _{K}$ was analyzed in Ref. 
\cite{PhysC}. For zero CMM, $K=0$, it becomes a single elementary integral,
with the familiar \cite{Coo} solution $\Delta _{0}=2\hbar \omega
_{D}/(e^{2/\lambda }-1)$ valid for {\it all} coupling $\lambda $. For small $%
K$ one determines \cite{PhysC} for weak coupling, $\lambda \rightarrow 0$,
that 
\begin{equation}
{\Delta _{K}}\mathrel{\mathop{\longrightarrow}\limits_{K \rightarrow 0}}%
\Delta _{0}-{\frac{2}{\pi }}\hbar v_{F}K+O(K^{2})  \label{eq:linear}
\end{equation}
where $v_{F}$ $\equiv \sqrt{2E_{F}/m}$ is the Fermi velocity. This {\it %
linear dispersion relation} is the 2D analog of the 3D result discussed as
far back as 1964 in Ref. \cite{Schrieffer}, p. 33 (see also, Ref. \cite{FW},
p. 336, and \cite{Adh}) but with the 2D coefficient $2/\pi$ in (\ref
{eq:linear}) replaced by $1/2$. Though commonly confused with the
Anderson-Bogoliubov-Higgs (ordinary) sound mode, the linear-dispersion
result (\ref{eq:linear}) corresponds to real, moving (pre-formed) CPs and is 
{\it distinct }from the zero-coupling ABH (indeed, IFG) phonons described by 
$\hbar v_{F}K/\sqrt{3}$. A general many-body formalism unambiguously
exhibiting \cite{Honolulu}\ this distinction involves solving the
Bethe-Salpeter equation for Cooper pairing based not on the IFG, as above,
but on the BCS ground-state in a Green's functions scheme allowing holes on
a par with particles.

For $N_{B}$ ordinary bosons of mass $m_{B}$ and energy $\varepsilon
_{K}=C_{s}\,K^{s}$ with $s$\ $>0$ and $C_{s}$ a constant, a temperature {\it %
singularity} appears at $T_{c}\neq 0$ for any \cite{Gunton}\ dimension $d>s$
in the number equation $N_{B}=\sum_{{\bf K}}[e^{(\varepsilon _{K}-\mu
_{B})/k_{B}T}-1]^{-1}$ at vanishing bosonic chemical potential $\mu _{B}\leq
0$ when the number of ${\bf K}=0$ bosons just ceases to be negligible upon
cooling. \ It is given \cite{cas} by 
\begin{equation}
T_{c}=\frac{C_{s}}{k_{B}}\left[ \frac{s\,\Gamma (d/2)\,(2\pi )^{d}n_{B}}{%
2\pi ^{d/2}\,\Gamma (d/s)g_{d/s}(1)}\right] ^{s/d}
\end{equation}
with $n_{B}$ $\equiv N_{B}/L^{d}$\ the boson particle density, and $%
g_{d/s}(z)$ the usual Bose integrals expressible as the series 
\begin{equation}
g_{\sigma }(z)=\sum_{l=1}^{\infty }\frac{z^{l}}{l^{\sigma }}%
\mathrel{\mathop{\longrightarrow}\limits_{ z \rightarrow 1}}\zeta (\sigma ),
\label{gsigma}
\end{equation}
where $\zeta (\sigma )$ the Riemann zeta function of order $\sigma $. The
last identification in (\ref{gsigma}) holds when $\sigma >1$ for which $%
\zeta (\sigma )<\infty $, while the series $g_{\sigma }(1)$ diverges for $%
\sigma \leq 1$, thus giving $T_{c}=0$ for $d\leq 2$. For $s=2$ and $d=3$ one
has $\zeta (3/2)\simeq 2.612$, and since $C_{2}\equiv $ $\hbar ^{2}/2m_{B}$
(3) then reduces to the familiar formula $T_{c}\simeq 3.31\hbar
^{2}n_{B}^{2/3}/m_{B}k_{B}$ of ``ordinary'' Bose-Einstein condensation
(BEC). On the other hand, for either particle {\it or}\ hole bosons with
(positive) excitation energy $\varepsilon _{K}\equiv \Delta _{0}-\Delta _{K}$
given asymptotically by the linear term in (\ref{eq:linear}) for all $K$, we
have $C_{1}\equiv a(d)\hbar v_{F}$ where \cite{Fujita1} $a(d)=1,\ 2/\pi $
and $1/2$ for $d=1,$ $2$ and $3$, respectively. Now $T{_{c}}$ is {\it nonzero%
} {\it for all} ${d}$ $>{1}$---which is {\it precisely} the dimensionality
range of all known superconductors including the quasi-1D organo-metallic
(Bechgaard) salts \cite{salts}.

The number of bosons in the boson-fermion mixture in chemical/thermal
equilibrium turn out \cite{physA}\ to be temperature-dependent, and it is 
{\it in conserving the fermion number} that the singularity arises. As in
the case of the pure boson gas (3), a linear rather than a quadratic
dispersion relation is needed to obtain BEC in {\it exactly}\ 2D. All this
emerges in a statistical model for the ideal binary gas {\it mixture} of
bosons (the CPs) and unpaired (both pairable and unpairable) fermions in
chemical equilibrium \cite{Schafroth}. \ Thermal pair-breaking of the bosons
into unpaired pairable fermions is explicitly allowed. At any $T$\ the total
number of fermions in 2D is $N=L^{2}k_{F}^{2}/2\pi =N_{1}+N_{2}$, and is
just the number of non-interacting (i.e., unpairable) fermions $N_{1}$ plus
the number of pairable ones $N_{2}$. The unpairable fermions obey the usual
Fermi-Dirac (FD) distribution with the IFG chemical potential $\mu $. On the
other hand, the $N_{2}$ pairable fermions are simply those in the
interaction shell of energy width $2\hbar \omega _{D}$ so that 
\begin{equation}
N_{2}=2\int_{\mu -\hbar \omega _{D}}^{\mu +\hbar \omega _{D}}\;d\epsilon 
\frac{g(\epsilon )}{e^{\beta (\epsilon -\mu )}+1}=2g\hbar \omega _{D},
\label{n214}
\end{equation}
where $\beta \equiv (k_{B}T)^{-1}$,\ since a constant $g(\epsilon )$ renders
the remaining integral exact. At any interfermionic coupling and temperature
these fermions form an ideal mixture of pairable but unpaired fermions plus
CPs that are created near the single-fermion energy $\mu (T)$, with binding
energy $\Delta _{K}(T)$ $\geq 0$ and total energy 
\begin{equation}
E_{K}(T)\equiv 2\mu (T)-\Delta _{K}(T).  \label{bosonenergy}
\end{equation}
This generalizes the $T=0$ definition $E_{K}\equiv 2E_{F}-\Delta _{K}$ given
below (1).

The Helmholtz free energy $F\equiv E-TS$, where $E$ is the internal energy
and $S$ the entropy, for this binary gas {\it ``composite
boson/pairable-but-unpaired-fermion system''} at $T\leq T_{c}$ is then
readily constructed \cite{physA} in terms of: a) $n_{2}(\epsilon )$, the
average number of unpaired but pairable fermions with energy $\epsilon $; b) 
$N_{B,0}(T)$, the number of (bosonic) CPs with zero CMM at temperature $T$;
and c) $N_{B,K}(T)$, that number of excited pre-formed CPs (i.e., with
arbitrary nonzero CMM $K$) and a cutoff $K_{0}$ physically defined \cite
{PhysC} by $\Delta _{K_{0}}\equiv 0$ denoting the value of $K$ beyond which
a CP breaks up. The free energy $F_{2}$ of the {\it pairable} fermions is to
be minimized subject to the constraint that $N_{2}$ is conserved. \ If $%
N_{20}(T)$ is the number of pairable but unpaired fermions, the relevant 
{\it number equation} for the pairable (i.e., active) fermions is then 
\begin{eqnarray}
N_{2} &=&N_{20}(T)+2[N_{B,0}(T)+N_{B,0<K<K_{0}}(T)]  \nonumber \\
&\equiv &N_{20}(T)+2N_{B}(T),
\end{eqnarray}
where $N_{B,0<K<K_{0}}(T)$ denotes the {\it total} number of ``excited''
bosonic pairs (namely with CMM such that $0<K<K_{0}$), i.e., $%
N_{B,0<K<K_{0}}(T)\equiv \sum_{0<K<K_{0}}N_{B,K}(T)$. At $T=0$ two distinct
coupling regimes emerge: a) for $\Delta _{0}<2\hbar \omega _{D}$ or for $\lambda \leq 2/\ln 2\simeq 2.89$, we have that $N_{20}(0)=g(2\hbar \omega
_{D}-\Delta _{0})$; while b) for $\Delta _{0}>2\hbar \omega _{D}$ (or $%
\lambda \geq 2.89$), $N_{20}(0)$ is identically zero. Hence, the number of
bosons $N_{B}(0)$ at $T=0$\ from (7) is just $N_{B}(0)={\frac12}[N_{2}-N_{20}(0)]$. Using (\ref{n214}) for $N_{2}$ the {\it 
fractional number of pairable fermions that are actually paired} at $T=0$,
namely $2N_{B}(0)/N_{2}=1-N_{20}(0)/N_{2}$, becomes simply $\Delta
_{0}/2\hbar \omega _{D}=(e^{2/\lambda }-1)^{-1}\mathrel{\mathop{%
\longrightarrow}}e^{-2/\lambda}$ as $\lambda \rightarrow 0$,\ for$\ \lambda
\leq 2/\ln 2\simeq 2.89$, and unity for $\lambda \geq 2/\ln 2$. As $N_{B}(0)=%
{\frac12}g\Delta _{0}$ for $\lambda \leq 2.89$, only those fermions in an
energy shell of width ${\frac12}\Delta _{0}$ around the Fermi surface
actually pair at $T=0$, while for $\lambda \geq 2.89$ {\it all} pairable
fermions pair up since then $N_{B}(0)=g\hbar \omega _{D}\equiv {\frac12}N_{2}
$. For $T>0$, $\ 2N_{B}(T)/N_{2}=1-N_{20}(T)/N_{2}$ {\it decreases }with $T$%
, provided one knows $\Delta _{0}(T)$ for any $T\geq 0$ and ascertains that
it decreases. For $T>0$, the $\theta (k_{1}-k_{F})\equiv \theta (\epsilon
_{k_{1}}-E_{F})$ 
in (\ref{eq:cooper}) becomes $1-n(\xi _{k_{1}})$, where $n(\xi
_{k_{1}})\equiv (e^{\beta \xi _{k_{1}}}+1)^{-1}$ is the FD distribution with 
$\xi _{k_{1}}\equiv \epsilon _{k_{1}}-\mu (T)$, with the IFG chemical
potential $\mu (T)$ in 2D given exactly by $\mu (T)=\beta ^{-1}\ln (e^{\beta
E_{F}}-1)\mathrel{\mathop {\longrightarrow
}}E_{F}$ as $T\rightarrow 0$.\ Similarly for $\theta (k_{2}-k_{F})$. Since $%
k_{1}=k_{2}$ implies that $\xi _{k_{1}}=\xi _{k_{2}}$, (\ref{eq:cooper})
then leads to a simple generalization to nonzero $T$\ of the $K=0$ CP
equation, 
\begin{equation}
1=\lambda \int_{0}^{\hbar \omega _{D}}d\xi (e^{-\beta \xi }+1)^{-2}[2\xi
+\Delta _{0}(T)]^{-1}.  \label{e13}
\end{equation}
Numerical solution shows $\Delta _{0}(T)$ to indeed be monotonic-decreasing
in $T$ for any fixed $\lambda $ and $\hbar \omega _{D}$. Further, the
solution of $\Delta _{0}(T^{\ast })=0$ is, by inspection, $T^{\ast }=\infty $%
; this infinite ``de-pairing'' temperature is unrealistic and undoubtedly an
artifact of the simplest version (1) of Cooper pairing used here as a
starting point; see, however, Ref. \cite{Honolulu}.

Modeling our system as a {\it pure boson gas} of CPs (i.e., neglecting the
background unpaired fermions) but with a temperature-dependent number
density $n_{B}(T)$ converts the explicit $T_{c}$-formula (3) into an {\it %
implicit} one. For $s=1$ and $d=2$ it becomes, since $g_{2}(1)\equiv \zeta
(2)=\pi ^{2}/6$, 
\begin{equation}
T_{c}=\frac{4\sqrt{3}}{\pi ^{3/2}}\frac{\hbar v_{F}}{k_{B}}\sqrt{n_{B}(T_{c})%
}.  \label{(11)}
\end{equation}
This requires $n_{B}(T)\equiv N_{B}(T)/L^{2}$ which in turn requires $\Delta
_{0}(T)$ as determined from (\ref{e13}), and follows from the expression $%
2N_{B}(T)/N_{2}=1-N_{20}(T)/N_{2}$. Solving this self-consistently with (7)
for $\lambda =1/2$ gives the remarkably constant value $T_{c}/T_{F}\simeq
0.004$, where $T_{F}\equiv E_{F}/k_{B}$, over the entire range of $\nu
\equiv \hbar \omega _{D}/E_{F}$ values $0.03-0.07$ typical of cuprate
superconductors. On the other hand, the BCS formula $T_{c}^{BCS}\simeq
1.13\Theta _{D}e^{-1/\lambda }$ with $\lambda =1/2$ gives $T_{c}/T_{F}$ $%
=0.005$ to $0.011$ over the same range of $\nu $ values. Obviously, both
sets of predictions are too small compared with the empirical cuprate range $%
T_{c}/T_{F}\simeq $ $0.03-0.09$ \cite{Poole}.

The exact $T_{c}$ {\it without} neglecting the background unpaired fermions
requires the exact CP excitation energy dispersion relation $\varepsilon
_{K}(T)\equiv \Delta _{0}(T)-\Delta _{K}(T)$ which is neither precisely
linear in $K$ nor independent of $T$. To determine $\Delta _{K}(T)$ we need
a working equation that generalizes Ref. \cite{PhysC} for $T>0$ via the new
CP eigenvalue equation (\ref{e13}). At $T=T_{c}$ both $N_{B,0}(T_{c})\simeq
0 $ and $\mu _{B}(T_{c})\simeq 0$ so that one gets \cite{physA} the implicit 
$T_{c}$-equation for the {\it binary gas mixture} 
\begin{eqnarray}
1 &=&{\frac{\tilde{T}_{c}}{\nu }}\ln \left[ {\frac{1+e^{-\{\tilde{\Delta}%
_{0}(\tilde{T}_{c})/2-\nu \}/\tilde{T}_{c}}}{1+e^{-\{\tilde{\Delta}_{0}(%
\tilde{T}_{c})/2+\nu \}/\tilde{T}_{c}}}}\right] +{\frac{8(1+\nu )}{\nu }} 
\nonumber \\
&&{\times }\int_{0}^{\kappa _{0}(\tilde{T}_{c})}d\kappa {\frac{\kappa }{e^{[%
\tilde{\Delta}_{0}(\tilde{T}_{c})-\tilde{\Delta}_{\kappa }(\tilde{T}_{c})]/%
\tilde{T}_{c}}-1}},  \label{Tcnum}
\end{eqnarray}
where quantities with tildes are in units of $E_{F}$ or $T_{F}$; $\kappa
\equiv K/2(k_{F}^{2}+k_{D}^{2})^{%
{\frac12}%
}$ with $k_{D}$ defined through $\hbar \omega _{D}\equiv \hbar
^{2}k_{D}^{2}/2m$; and $\nu \equiv \Theta _{D}/T_{F}$.\ To obtain $T_{c}$
from the finite-$T$ dispersion relation one must numerically\ solve {\it four%
} equations self-consistently for each $\lambda $ and $\nu $, namely (\ref
{Tcnum}) in conjunction with (8) for $\tilde{\Delta}_{0}(\tilde{T})$, and
Eq. (35) of \ Ref. \cite{PhysC} for both $\tilde{\Delta}_{\kappa }(\tilde{T}%
) $ and the breakup value $\kappa _{0}(\tilde{T}_{c})$. For $\lambda =1/2$
and the range of $\nu $ values $0.03-0.07$ typical of cuprates, the
resulting $T_{c}/T_{F}$ falls within the aforementioned empirical range $%
0.03-0.09$ \cite{Poole}. \ For cuprates $d\simeq 2.03$ has been suggested 
\cite{wen} as more realistic since it reflects inter-CuO-layer couplings,
but our results in that case would be very close to those for $d=2$ since,
e.g., from (3) $T_{c}$ for $s=1$ (but {\it not }for $2$)\ varies little with 
$d$ around $d=2$. In fact, if $m_{B\perp }$ and $m_{B}$\ are\ the boson
masses {\it perpendicular} and {\it parallel}, respectively,{\it \ }to the
cuprate planes, an ``anisotropy ratio'' $m_{B}/m_{B\perp }$ varied from 0 to
1 allows ``tuning'' $d$ continuously from 2 to 3.

Other boson-fermion models \cite{Rann85}\cite{TDLee}\cite{Gesh}\cite{Levin} 
\cite{Domanski} have been introduced, some even addressing \cite{Gesh}\cite
{Levin}\ $d$-wave interaction effects as opposed to the pure $s$-wave
considered here, and some also focus\ \cite{Levin}\cite{Domanski}\ on the
pseudogap. \ But calculating cuprate $T_{c}$ values in quasi-2D without
adjustable parameters has not been attempted, and indeed $T_{c}\equiv 0$ is
predicted in exactly 2D.

In summary, a simple statistical model treating pre-formed CPs with {\it both%
} zero and nonzero CMM as non-interacting bosons in chemical/thermal
equilibrium with unpaired fermions is proposed that gives rise to a boson
number that is strongly coupling- and temperature-dependent. Since the CP
dispersion relation is approximately linear for nonzero CMM, it exhibits a
BEC of zero-CMM pairs at precisely 2D. In contrast to both BCS
theory---which is {\it not} \cite{Bardeen}{\it \ }a BEC theory---and simpler
BE models excluding either the breakable character of the pre-formed CPs or
the presence of unpaired fermions, exact $T_{c}$'s for the boson-fermion
binary mixture based upon the {\it exact }CP dispersion relation are found
in rough agreement with empirical cuprate values with no adjustable
parameters.

M.C., A.P. and A.R. are grateful for partial support from grant PB98-0124,
and M.deLl. from grants PB92-1083 and SAB95-0312, both by DGICYT (Spain),
and from PAPIIT IN102198-9 as well as CONACyT 27828-E (both Mexico). \
M.deLl. thanks P.W. Anderson, D.M. Eagles, R. Escudero, S. Fujita, M.
Fortes, K. Levin, O. Rojo, A.A. Valladares for discussions, and V.V.
Tolmachev for correspondence as well.

\end{document}